\documentclass[pra, twocolumn, showpacs, preprintnumbers, amsmath, amssymb, superscriptaddress, aps]{revtex4}
\usepackage{amssymb}
\usepackage{latexsym}
\usepackage[dvips]{graphicx}

\usepackage{epsfig}

\usepackage{dcolumn}
\usepackage{amsmath}
\usepackage{amsfonts}
\usepackage{bm}
\usepackage{color}

\newcommand{\be}{\begin{equation}}
\newcommand{\ee}{\end{equation}}
\newcommand{\bea}{\begin{eqnarray}}
\newcommand{\eea}{\end{eqnarray}}

\newcommand{\p}{\partial}
\newcommand{\s}{\sigma}

\newcommand{\la}{\langle}
\newcommand{\ra}{\rangle}
\newcommand{\rd}{\mbox{d}}
\newcommand{\ri}{\mbox{i}}
\newcommand{\re}{\mbox{e}}

\renewcommand{\vec}[1]{{\bm #1}}

\begin{document}
\title{The ladder physics in the Spin Fermion model}
\author{A. M. Tsvelik}
\affiliation{Condensed Matter Physics and Materials Science Division, Brookhaven National Laboratory, Upton, NY 11973-5000, USA}
 \date{\today } \begin{abstract} 
A link is established between the spin-fermion (SF) model of the cuprates and the approach based on the analogy between the physics of doped Mott insulators in two dimensions and the physics of fermionic ladders. This enables one  to use nonperturbative results derived for fermionic ladders to move beyond the large-N approximation in the SF model. It is shown that the paramagnon exchange postulated in the SF model has exactly the right form to facilitate the emergence of the fully gapped d-Mott state in the region of the Brillouin zone at the hot spots of the Fermi surface. Hence the SF model provides an adequate description of the pseudogap. 
 
\end{abstract}


\maketitle

 {\bf Introduction} The purpose of this letter is (i) to establish a link between  two successful phenomenological descriptions of the underdoped cuprates and (ii) to use  it to obtain some new results. One description is based on the spin-fermion (SF) model \cite{SF}; the other one is based on the analogy between the cuprate physics and physics of fermionic ladders.  Both approaches have  been successful in describing different aspects of the cuprate physics. The SF model provides a natural platform for  description of the complicated phase diagram of the cuprates (see, for example, \cite{wang} and references therein). The merit of the ladder description, on the other hand, lays in its treatment of the pseudogap. In this approach the formation of  pseudogap does not require any translational symmetry breaking which is consistent with  the experiments \cite{Ex1,Ex2,Ex3}. The ladder analogy  was first put forward  by Dagotto and Rice \cite{dagotto};   the insights from models of coupled ladders \cite{krt} had eventually lead to the popular  expression for the  single electron Green's function (the so-called YRZ expression) \cite{yrz}.  

The SF model describes electrons interacting with soft paramagnons in  a vicinity of the antiferromagnetic Quantum Critical Point (AF QCP). The strongest correlations are concentrated in the regions of Fermi surface connected by the antiferromagnetic wave vectors ${\bf Q} = (\pm \pi,\pm \pi)$ (hot spots).  The low energy degrees of freedom  are collective magnetic excitations and quasiparticles with a large Fermi surface. The model  was conceived as phenomenological, but similar descriptions can be obtained from microscopic models such as the  model of weakly coupled Hubbard ladders \cite{essler} or  two dimensional Hubbard model  using  Dynamical Mean Field Theory (DMFT) \cite{xu}.  Below I will demonstrate that  the ladder physics is contained in the SF model. The effective dimensionality reduction related to the ladder physics is facilitated by  the long range exchange interaction mediated by the paramagnons. 

 The Hamiltonian of the SF model is 
  \bea
 && H = \sum_k \epsilon(k)\psi^+_{\s}(k)\psi_{\s}(k) + g\sum_{k,q}\psi^+(k+q)\vec\s\psi(k){\bf S}(q) +\nonumber\\
 && \sum_q \frac{1}{2} {\bf S}(-q)\chi^{-1}(q){\bf S}(q), \label{SF}\\
 && \chi^{-1}(q) = \chi_0^{-1}[\xi^2({\bf Q}-{\bf q})^2 + 1].\nonumber
 \eea
 It has been shown that the frequency dependence of the spin susceptibility $\chi$ is generated by the quasiparticles and does not need to be included in the bare action \cite{SF}. 
 
  In the standard approach the spin-fermion coupling is considered as a perturbation which generates  significant effects  only near  the spots on the Fermi surface connected by ${\bf Q}$-vector. The pieces of FS connected by the ${\bf Q}$-vectors are not nested.  
 For the most recent applications of this approach  one can consult \cite{wang}. 
  
 {\bf Ladder model approach}. I this letter I will follow  a different approach. I will assume that the spin-fermion coupling is sufficiently strong and take as a starting point the FS which is nested at the hot spots. If the spectral gaps generated by the interaction are sufficiently large then one can treat the deviation from the nesting  as a perturbation. One can also speculate that the strong interactions will modify the shape of the Fermi surface to stabilize nesting to take advantage of the gap opening, as it happens in the commensurate antiferromagnetic state in Cr alloys \cite{riceAF} (see also \cite{rice}).  

 As I have stated above, due to the singular character of the spin-spin interaction the strong correlations occur only in the vicinity of hot spots. The eight spots are divided in two quartets; in one quartet the spots from the opposite sides of the FS are connected by the wave vector $(\pi,\pi)$ and in the other by the wave vector $(\pi,-\pi)$.   To the first approximation these two groups of hot spots can be considered independently.
 
 \begin{figure}
\centerline{\includegraphics[angle = 0,
width=1.5\columnwidth]{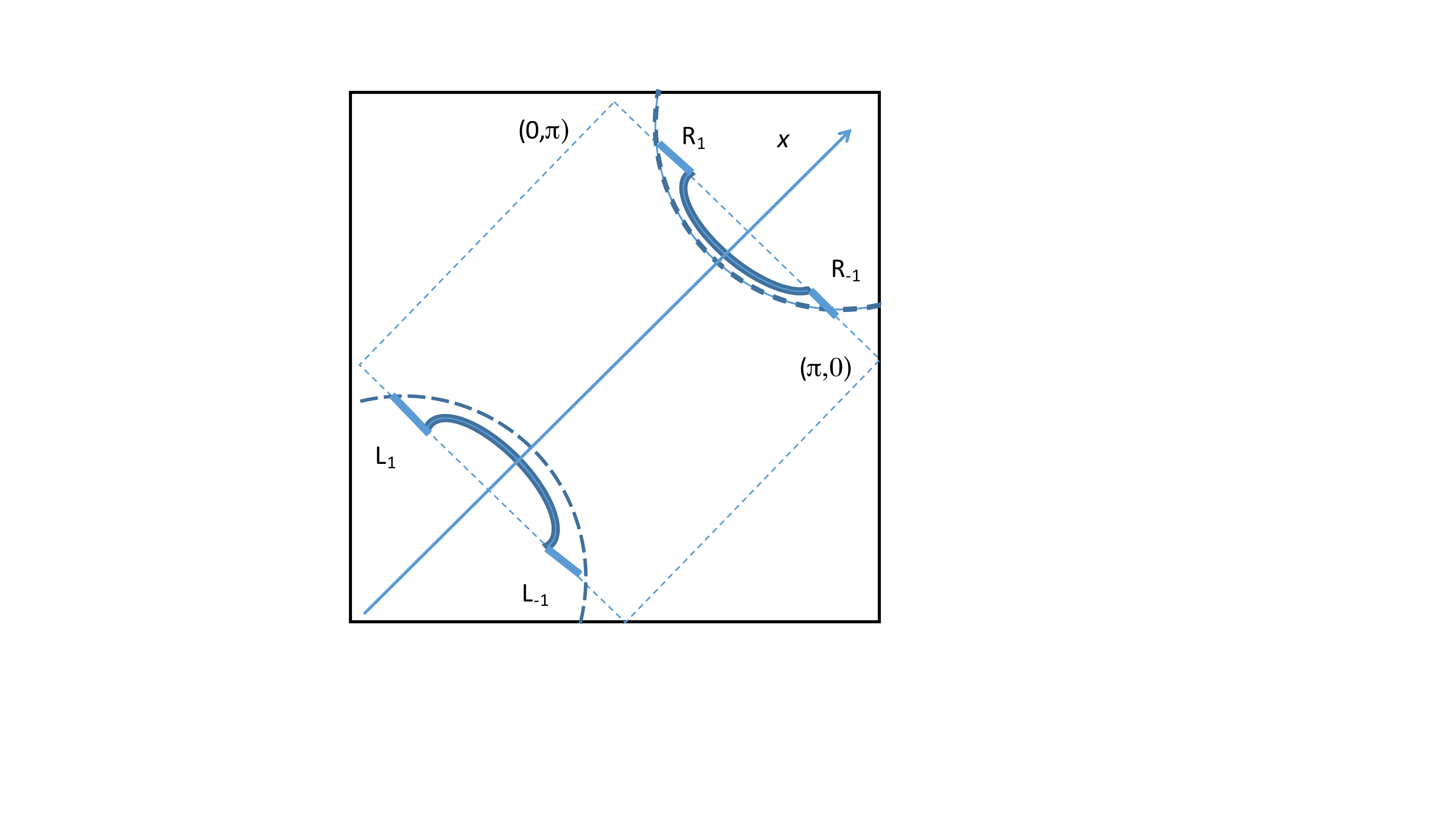}}
\caption{A sketch of a part of the Fermi surface with four hot spots. The dashed curves correspond to the Fermi surface of noninteracting electrons. It is energetically advantageous for  the interactions to deform the Fermi surface  making it flat at the hot spots and facilitating nesting of the opposite parts of the FS (drawn by the solid lines). The right (left) moving fermions are located near the flat portions of the Fermi surface and are marked by $R_1,R_{-1}$ ($L_1,L_{-1}$). Hot spots with the same number are connected by wave vector $(\pi,\pi)$. The ones connected by $(\pi,-\pi)$ are not shown. The line with an arrow  corresponds to the coordinate axis of $x$. }
 \label{pict1}
\end{figure}
 
 It is convenient to use a tomographic projection and to introduce operators 
 \bea
 && R_a(x) = \frac{1}{\sqrt{2\pi}}\int \psi({\bf k}_R^a +k_{\parallel}{\bf e})\re^{\ri k_{\parallel}x}, \nonumber\\
 &&  L_a(x) = \frac{1}{\sqrt{2\pi}}\int \psi({\bf k}_L^a +k_{\parallel}{\bf e})\re^{\ri k_{\parallel}x},
 \eea
 where ${\bf k}^a_{R,L}$ $a=\pm 1$ are coordinates of the hot sports in momentum space and ${\bf e} = (1,1)/\sqrt 2$. Then after integration over the spin variables, we obtain the following Hamiltonian density, describing the vicinity of hot spots:   
   \bea
   && {\cal H} = \ri v_F\sum_{a=\pm 1,\s = \pm 1/2}(-R^+_{a\s}\p_x R_{a\s} + L^+_{a\s}\p_x L_{a\s}) +V, \nonumber\\
  && V = -\gamma \sum_{a=\pm 1} (R^+_a\vec\s L_a + L^+_a\vec\s R_a)\sum_{b=\pm 1}(R^+_b\vec\s L_b+ L^+_b\vec\s R_b), \nonumber\\
  && \gamma = rg^2\chi_0, ~~ r \sim \xi^{-1}. \label{GNV}
  \eea 
 There is a similar Hamiltonian for the quartet of hot spots connected  by the wave vector $(\pi,-\pi)$. To the first approximation these two groups of hot spots can be considered independently.
 
    The form of the interaction (\ref{GNV}) has the form  used in  the models of half filled ladders. These models have been thoroughly studied using nonperturbative methods. Hence we can go beyond the standard large-$N$ ($N$ is the number of hot spots) approach. The symmetry of the model is $U(1)\times U(1)\times SU(2)\times Z_2$. The $U(1)$ symmetries refer to the charge conservation inside of every patch, the $SU(2)$ symmetry refers to spin and the $Z_2$ symmetry refers to the fact that the patches can be interchanged. The RG calculations demonstrate \cite{balents, lin}  that this original symmetry increases dynamically when the model scales to strong coupling such that at low energies  the maximally allowed symmetry, namely, the O(8) symmetry emerges. The O(8)-symmetric Gross-Neveu model is exactly solvable \cite{GN}, which allows one to extract a great deal of information. When the bare  coupling $\gamma$ is not small  the symmetry becomes approximate, but many statements remain valid on a qualitative level \cite{tsvelik}.  It is particularly important for us that, as was established in \cite{lin},\cite{ludwig},\cite{azaria}, that the symmetry itself  does not uniquely fix the ground state properties of the model. This is related to the fact that there are transformations of the Hamiltonian which do not change the excitation spectrum, but change the observables. In \cite{azaria} it was established that these transformations are automorphisms of the O(8) group. As a consequence, the phase diagram of the ladder includes different phases, some of them are favorable for superconductivity, and some are not (see the discussion in \cite{controzzi} and in Supplemental Material). To find out to what part of the phase diagram the interaction scales requires certain analysis. 
    
    The interaction (\ref{GNV}) is divided into two parts: inside of a hot spot (labeled by indices $\pm 1$) and between the spots. For the former ones we have 
    \bea
    &&V_{11} + V_{-1,-1} = \\
    && \gamma \sum_{a=\pm 1}\Big[ L^+_{\s}L_{\s} R^+_{\s'}R_{\s'} + 3(R^+_{\s} R^+_{\s'}L_{\s'}L_{\s} + H.c.) \nonumber\\
    &&- (L^+\vec\s L)(R^+\vec\s R)\Big]_a\nonumber
   \eea
   This part of the interaction contains the umklapp processes which are responsible for the Mott physics. The interaction between the spots looks like the interchain exchange in the ladder model with no interchain tunneling, studied in \cite{shelton}:
   \be
   V_{1,-1} = -\gamma \int \rd x {\bf S}_1(x){\bf S}_{-1}(x) \label{V12}
   \ee
   We  bosonize the model using the standard notations:
   \bea
   && R_{p\s} = \frac{\eta_{p\s}}{\sqrt{2\pi a_0}}\exp[-\ri(\varphi_c +p\varphi_f + \s\varphi_s +p\s\varphi_{sf})], \nonumber\\
   &&  L_{p\s} = \frac{\eta_{p\s}}{\sqrt{2\pi a_0}}\exp[\ri(\bar\varphi_c +p\bar\varphi_f + \s\bar\varphi_s +p\s\bar\varphi_{sf})], \label{ferm}
  \eea
  where $a_0$ is the lattice constant,  $p=\pm 1, \s = \pm 1$,  and 
  \be
  \{\eta_a,\eta_b\} = 2\delta_{ab},
  \ee
  are Klein factors, and $\varphi_a,\bar\varphi_a$ are chiral bosonic fields with commutation relations $[\varphi(x),\varphi(y)] = -(\ri/4)$sign$(x-y)$, $[\bar\varphi(x),\bar\varphi(y)] = (\ri/4)$sign$(x-y)$, $[\varphi(x),\bar\varphi(y)] = -\ri/4$. 
   Then we have 
   \bea
   V_{1,-1} = \frac{\ri\gamma}{\pi a_0}(\cos\sqrt{4\pi}\Phi_c + \cos\sqrt{4\pi}\Phi_f)(\vec\xi_R\vec\xi_L - 3\xi_R^0\xi_L^0) \label{V1}
   \eea
   The easiest way to see the difference between the phases is to rewrite the model in terms of Majorana fermions (see Supplementary Material for details). The ones in the spin sector are defined as 
\bea
&& \xi_R^1 = \xi\frac{\cos(\sqrt{4\pi}\varphi_s)}{\sqrt{2\pi a_0}}, ~~ \xi_R^2 = \xi\frac{\sin(\sqrt{4\pi}\varphi_s)}{\sqrt{2\pi a_0}}, \nonumber\\
&& \xi_R^3 = \eta\frac{\cos(\sqrt{4\pi}\varphi_{sf})}{\sqrt{2\pi a_0}}, ~~ \xi_R^0 = \eta\frac{\sin(\sqrt{4\pi}\varphi_{sf})}{\sqrt{2\pi a_0}}, \label{Maj}
\eea
where $\{\xi,\eta\} =0, \xi^2=\eta^2 =1$ are Klein factors. Together with the Majoranas from the charge sector which are made from chiral components of $\Phi_c,\Phi_f$, they comprise an octet of Majorana fermions transforming according to the vector representation of the O(8) group.  The original  fermions and their antiparticles transform according to irreducible spinor representations of the group.  As is clear from comparison (\ref{ferm}) and (\ref{Maj}), the two groups of  the fermions  are {\it nonlocal} with respect to each other. However, the  introduction of the Majoranas simplifies the Hamiltonian reducing it to the Gross-Neveu model form (we set $v_F=1$): 
 
  \bea
  && {\cal L} = \sum_{a=c,f}(1+\gamma/\pi)(\p_{\mu}\Phi_a)^2 + V + \label{V}\\
  && \frac{1}{2}\sum_{a=0}^3[\xi_R^a(\p_{\tau} -\ri \p_x)\xi_R^a +\xi_L^a(\p_{\tau} + \ri\p_x)\xi_L^a],\nonumber\\
  && V = \gamma\Big[ -\frac{3}{2(\pi a_0)^2} \cos(\sqrt{4\pi}\Phi_c)\cos(\sqrt{4\pi}\Phi_f) + \nonumber\\
  && \frac{\ri}{(\pi a_0)} (\cos\sqrt{4\pi}\Phi_c + \cos\sqrt{4\pi}\Phi_f)(\vec\xi_R\vec\xi_L - 3\xi_R^0\xi_L^0) \nonumber\\
&& -(\vec\xi_R\vec\xi_L+\xi_R^0\xi_L^0)^2\Big], \nonumber
  \eea
Here I decided to leave the charge sector in the bosonic form which makes it easier to operate with the order parameters.  This form is manifestly $U(1)\times U(1)\times SU(2)\times Z_2$-symmetric. Two bosonic fields $\Phi_c,\Phi_f$ describe the charge sector. The spin sector is described by the four Majorana fermions, three of which transform according to the triplet and one according to the singlet representations of the SU(2) group. The $Z_2$ symmetry is realized as the invariance of the Lagrangian under a sign change of all  Majorana fermions.  The renormalization group calculations indicate that the interaction scales to strong coupling where it becomes O(8)-symmetric. During the flow the coupling at the last  term changes sign. The most important fact is that the signs of the cross terms in the interaction, the ones which include $M_{s}=\ri(\vec\xi_R\vec\xi_L)$ and $M_{sf}=\ri\xi_R^0\xi_L^0$ do not change under RG. This is because these signs determine the character of the ground state, as explained below (see \cite{controzzi} and Supplementary Material). 
  
 {\bf The order parameter}. Below I will argue that  the paramagnon exchange postulated by the SF model drives the system exactly into the ground state of the d-Mott phase \cite{lin}. This is important since this phase is "pregnant" with d-wave superconductivity and it turns out that the SF model has the right form of the interaction for this. Indeed,  the corresponding order parameter (OP) 
  \bea
  && \Delta_d = R_{1\uparrow}L_{2\downarrow} - R_{2\uparrow}L_{1\downarrow} - R_{1\downarrow}L_{2\uparrow} + R_{2\downarrow}L_{1\uparrow} =  A\re^{\ri\sqrt{\pi}\Theta_c}, \nonumber\\
&& A = \Big[-\cos(\sqrt\pi\Phi_f)\s_1\s_2\s_3\mu_0 +\nonumber\\
&& \ri\sin(\sqrt{\pi}\Phi_f)\mu_1\mu_2\mu_3\s_0\Big], \label{delta}
  \eea
  has a finite amplitude. Indeed, this OP can be conveniently factorized into the exponent of the dual charge field $\Theta_c$, which average is always zero in the Mott phase, and the amplitude $A$. The amplitude  has a vacuum average - that is what I mean by d-Mott phase being pregnant with the superconductivity. The latter one will emerge when the chemical potential will exceed a half of the Cooperon gap so that the field $\Theta_c$ becomes gapless. To make sure that the above description is correct, let us consider the amplitude closer. From (\ref{V}) we see that the vacuum of the theory corresponds to $\Phi_c=\Phi_f = 0$ or $\Phi_c = \Phi_f = \sqrt{\pi}/2$. The mass of Majorana fermions changes its sign when the fields interpolate from one vacuum to another. In the first vacuum we have $\la\s_a\ra,~(a=1,2,3), \la\mu_0\ra \neq 0$, in the second one $\la\mu_a\ra~(a=1,2,3), \la\s_0\ra \neq 0$ (see Supplementary Material). Hence the vacuum average of the amplitude is never zero.
  
   Now recall that the Gross-Neveu model (\ref{V})  scales to strong coupling and its spectrum is entirely gapped. The exact solution of the O(8)-symmetric model yields the spectrum which contains quasiparticles with quantum numbers of the electron (charge $\pm e$, spin $\pm 1/2$ and chain index $\pm 1$) and eight Majorana fermions transforming in the vector representation of the group. Among the latter ones are particle-particle bound states with momentum zero and charge $\pm 2e$ (Cooperons). According to Konik and Ludwig \cite{ludwig}, the correlation function of the OPs (\ref{delta}) in the d-Mott phase contains a pole, so the Cooperons  are coherent excitations. There are also  particle-hole bound states (SDW) which carry momentum $(\pi,\pi)$. 
  
   The spectral gaps in the ladder are formed as  a consequence of discrete symmetry breaking which occurs at $T=0$. In the d-Mott phase this symmetry is not a traslational one. The quasiparticle excitations are kinks, which interpolate between different degenerate ground states. At finite temperature the density of kinks is  finite  and  the spectral gaps are gradually filled by the thermal fluctuations. This picture corresponds to a gradual filling of the pseudogap, observed in ARPES  experiments such as \cite{Ex1}.  

{\bf Excitations}. The correlation functions of the d-Mott phase were studied  by Konik and Ludwig \cite{ludwig}. For instance, the single particle Green's function is  
 \bea
 G_q = \frac{\ri\omega + \epsilon_q(k)}{\omega^2 + \epsilon_q^2(k) + \Delta^2(q)}, \label{G}
 \eea
 where $k$ is perpendicular and $q$ is along the FS. The perturbation (the deviation of the actual FS from the flat one)  is just a $q$-dependent chemical potential, it must be added to $\ri\omega$: $\ri\omega \rightarrow \ri\omega + \delta\epsilon(q,k)$.  This makes the resulting Green's function to look closer to the YRZ form, although not quite. The applicability of our ladder approximation is restricted by the region of the FS where the chemical potential lies inside the gap, although this statement requires a correction.
 
 As I pointed out above, alongside the single-particle excitations which spectra are determined by the poles of (\ref{G}), the GN model also has collective excitations transforming according to the vector representation of the O(8) group. One of this excitations is Cooperon, a bound state of two electrons (holes) with momentum 0. Its spectrum lies below the two-particle continuum and therefore the Cooperon energy will vanish before the single particle gap is closed. This means that within the current scenario the superconductivity originates from the areas of momentum space close to the tips of the FS pockets. The mechanism is Cooperon condensation, as was suggested in \cite{krt}. Another excitation (also gapped) is a coherent spin-1 magnetic excitation centered at ($\pi,\pi$) which can be identified with the upper part of the ``hourglass'' spectrum of magnetic excitations (see, for instance \cite{hourglass}). 

 Let me summarize.  By adopting a hypothesis that the interactions modify the Fermi surface around the hot spots to generate nesting, I demonstrated that the spin-fermion model describes the  d-Mott phase physics near the hot spots. This result brings together two successful phenomenological approaches. It also  explains the mechanism of the pseudogap formation: the Fermi surface near the hot spots is truncated without translational invariance breaking. 
 
 I am grateful to A. Chubukov,  G. Kotliar, R. M. Konik, P. D. Johnson for interesting discussions and encouragement. My special thanks are to T. M. Rice who gave me the idea to use nested Fermi surface. The work was  supported by the U.S. Department of Energy (DOE),  Division  of Condensed Matter Physics and  Materials Science, under Contract No. DE-AC02-98CH10886.


  
  \section{Supplementary Material}
  
 {\bf Phases of fermionic ladder}. The fermionic ladder models include four species of charged fermions. Following the previous work I use the approach where the original fermions  are bosonized and then refermionized into new (Majorana) fermions as explained in the main text. The most general form of the interaction compatible with the $U(1)\times U(1)\times SU(2)\times Z_2$ symmetry of the ladder problem is 
  \bea
&&  V = \frac{1}{2} g_{ab}M_aM_b, ~~ g_{ab}= g_{ba}, \\
&& M_c =\frac{1}{\pi a_0}\cos(\sqrt{4\pi}\Phi_c) = \ri(r_1l_1+r_2l_2), \nonumber\\
&& M_f = \frac{1}{\pi a_0}\cos(\sqrt{4\pi}\Phi_f) = \ri(r_3l_3+r_4l_4),\nonumber\\
&& M_s = \ri(\vec\xi_R\vec\xi_L), ~~ M_{sf} = \ri\xi_R^0\xi_L^0.\nonumber
\eea
 Interaction (\ref{V}) is a particular case of this general one. In the strong coupling regime the diagonal couplings $g_{aa} <0$, as far as the off-diagonal matrix elements are concerned, their signs are not fixed because one can always change signs of $M_a$ by changing a sign of the Majorana fermion of one chirality, for instance
 \bea
 \xi^0_R \rightarrow -\xi^0_R, ~~ \xi_L^0 \rightarrow \xi_L^0, ~~ M_{sf} \rightarrow - M_{sf}
 \eea
  The O(8) symmetry requires all off-diagonal matrix elements to be equal modulo their sign and to be equal to the diagonal matrix elements:
  \be
  g_{aa} = g_{bb} = -|g_{ab}| <0, ~~ a\neq b.
  \ee
  The sign  ambiguity leaves room for different phases. Although signs of $g_{ab}$ ($a\neq b$) do not affect the excitation spectrum, they affect the observables. To be precise they affects those observables  which are nonlocal in terms of the Majoranas. One example is given in the main text (\ref{delta}). This OP is expressed in terms of order and disorder parameters of the quantum Ising models. We will use the following identities derived in \cite{Itzykson}:
  \bea
  && \sin(\sqrt\pi\Phi_s) \sim \s_1\s_2, ~~ \cos(\sqrt\pi\Phi_s) \sim \mu_1\mu_2, \nonumber\\
  && \sin(\sqrt\pi\Theta_s) \sim \mu_1\s_2, ~~ \cos(\sqrt\pi\Theta_s) \sim \s_1\mu_2, \nonumber\\
  && \sin(\sqrt\pi\Phi_{sf}) \sim \s_3\s_0, ~~ \cos(\sqrt\pi\Phi_{sf}) \sim \mu_3\mu_0, \nonumber\\
  && \sin(\sqrt\pi\Theta_{sf}) \sim \mu_3\s_0, ~~ \cos(\sqrt\pi\Theta_{sf}) \sim \s_3\mu_0. 
  \eea

From these identities it follows that the vacuum with positive Majorana mass corresponds to $\la\s\ra \neq 0, \la\mu\ra =0$ and the one with negative mass corresponds to $\la\s\ra =0, \la\mu\ra \neq 0$. 

 It is instructive to see what kind of other phases the  fermionic ladder may have. As it was established already in \cite{Lin} (see also \cite{Konik}, \cite{Davide}) there are  four  local order parameters (OPs) bilinear in  the bare fermions including (\ref{delta}). There is s-CDW (Charge Density Wave):
\bea
&& \Delta^s_{CDW} =  R^+_{1\s}L_{1\s} + R^+_{2\s}L_{2\s} = \\
&&  \re^{\ri\sqrt\pi\Phi_c}\Big[\cos(\sqrt\pi\Phi_f)  \mu_1\mu_2\mu_3\mu_0 + \ri \sin(\sqrt\pi\Phi_f)\s_1\s_2\s_3\s_0\Big].\nonumber
\eea
For this OP to  condense one needs to change  the sign of coupling at the term containing $(\vec\xi_R\vec\xi_L)$ in (\ref{V1}).
Another one is d-CDW:
\bea
&& \Delta^d_{CDW} =  R^+_{1\s}L_{1\s} - R^+_{2\s}L_{2\s} = \\
&& \re^{\ri\sqrt\pi\Phi_c}\Big[\sin(\sqrt\pi\Phi_f)  \mu_1\mu_2\mu_3\mu_0 + \ri \cos(\sqrt\pi\Phi_f)\s_1\s_2\s_3\s_0\Big].\nonumber
\eea 
This one condenses if one changes sign at $\xi_R^0\xi_L^0$ in (\ref{V1}). The last OP is the s-wave superconducting one
\bea
&& \Delta_s = R_{1\uparrow}L_{2\downarrow} + R_{2\uparrow}L_{1\downarrow}  -R_{1\downarrow}L_{2\uparrow} - R_{2\downarrow}L_{1\uparrow} = \nonumber\\
&&  A_s\re^{\ri\sqrt{\pi}\Theta_c}, \\
&& A_s = \Big[-\ri\sin(\sqrt{\pi}\Phi_{f})\s_1\s_2\s_3\mu_0  + \cos(\sqrt{\pi}\Phi_{f})\mu_1\mu_2\mu_3\s_0\Big].\nonumber
\eea
 This one condenses if one changes the sign of both $(\vec\xi_R\vec\xi_L)$ and $\xi^0_R\xi_L^0$ terms in (\ref{V1}).
 
\end{document}